\documentclass[12pt,aps,pra,preprint]{revtex4}
\usepackage{graphicx}
\usepackage{amsmath,amssymb,amsfonts}
\usepackage{natbib}

\newcommand{\be}{\begin{equation}}
\newcommand{\ee}{\end{equation}}

\begin{document}

\title{Interplay of causticity and vorticality within the complex
quantum Hamilton-Jacobi formalism}

\author{A. S. Sanz}
\email{asanz@imaff.cfmac.csic.es}

\author{S. Miret-Art\'es}
\email{s.miret@imaff.cfmac.csic.es}

\affiliation{Instituto de F\'{\i}sica Fundamental,\\
Consejo Superior de Investigaciones Cient\'{\i}ficas,\\
Serrano 123, 28006 Madrid, Spain}

\date{\today}

\begin{abstract}
Interference dynamics is analyzed in the light of the complex quantum
Hamilton-Jacobi formalism, using as a working model the collision of
two Gaussian wave packets.
Though simple, this model nicely shows that interference in quantum
scattering processes gives rise to rich dynamics and trajectory
topologies in the complex plane, both ruled by two types of
singularities: caustics and vortices, where the former are associated
with the regime of free wave-packet propagation, and the latter with
the collision (interference) process.
Furthermore, an unambiguous picture connecting the complex and real
frameworks is also provided and discussed.
\end{abstract}



\maketitle


Realistic simulations of many quantum processes and phenomena of
interest ---e.g., diffusion, relaxation, transport, dephasing or
decoherence in solid state physics, condensed matter or chemical
physics--- require a detailed knowledge of the time evolution
(dynamics) of a relative large number of degrees of freedom.
A full quantum-mechanical study of this type of many-body problems via
the time-dependent Schr\"odinger equation (TDSE) results prohibitive
computationally.
Because of this inconvenience, this kind of problems has become an
important and challenging issue in the last years.
In particular, many efforts are being devoted to the development of
a number of alternative trajectory-based formalisms \cite{micha-bk}.
The most recent approach considered in the literature, in this
direction, is based on using the complex quantum Hamilton-Jacobi
(CQHJ) equation, formerly derived by Pauli \cite{pauli} in 1933 and
later on rediscovered by other authors \cite{leacock1,leacock2,hhl,%
sanz-jpcm1,sanz-jpcm2,john,tannor-bk}.
At a fundamental level, this formulation has been used basically to
describe the dynamics associated with stationary states \cite{john,%
yang1,yang2,yang3}, while time-dependent problems have received little
attention \cite{kabos}.
However, from a practical (numerical) viewpoint, it has received much
more attention \cite{leacock1,leacock2,hhl,wyatt11,wyatt12,wyatt13,%
wyatt21,wyatt22,tannor11,tannor12,tannor13,tannor2,poirier1}.
At the moment, the CQHJ equation has been applied to both
time-independent (bound states) and time-dependent (scattering)
problems, and is actually considered as a potential computational tool
to handle relatively large systems.
Note that, as also happens with classical waves and fields, solving
quantum problems within a complex framework is usually simpler than
in a real one.

One of the important problems in quantum trajectory-based methodologies
is that of interference dynamics.
This characteristic quantum-mechanical process, which is central to
many actual research fields in physics and chemistry (e.g., quantum
control \cite{paul-bk} and quantum information \cite{nielsen}),
constitutes however a numerical drawback for such techniques
\cite{tannor11,tannor12,tannor13,tannor2,wyatt-bk}, since it introduces
the so-called {\it nodal problem} \cite{wyatt-bk}: the nodal structures
associated with interfering amplitudes give rise to numerical
instabilities in the calculation of quantum trajectories (and the
properties derived from them).
In the literature, this issue has been tackled by means of different
strategies (see, for instance, Refs.~\cite{tannor2,nodal1,nodal2}).
One of them makes use of the superposition principle, separating the
contributions from each partial wave and then taking into account the
combined effects of all the contributing partial waves in the end.
Nevertheless, despite this strategy may result efficient numerically,
from a dynamical viewpoint (i.e., in terms of trajectories) it leads to
dramatic consequences: the trajectories associated with a superposition
look totally different to those associated with each separate partial
wave \cite{sanz-newepjd} due to quantum nonlocality \cite{sanz-cpl}.
The purpose of this Letter is to shed some light on this numerical
issue by analyzing the topology of complex quantum trajectories and
comparing them with their real homologous.
The unfolding in the complex space turns relatively simple dynamics
in real space into very intricate complex ones, where unexpected and
surprising features even for low-dimensional systems are observed.
A {\it vortical dynamics} in the complex configuration space appears
as a natural consequence of interference even in 1D, breaking the
{\it causticity regime} characterizing free propagation.
Two types of quantum singularities are thus shown to rule the complex
dynamics: vortices and caustics.
Apart from their intrinsic physical interest, these singularities
should also be taken into account when using this formulation for
computational purposes, in the design and implementation of numerical
algorithms.

For the sake of simplicity, here we consider the 1D CQHJ formulation,
although the results can be straightforwardly extended to higher
dimensions.
Thus, after considering the transformation relation
$\Psi (x,t) = e^{i\mathcal{S}(x,t)/\hbar}$, where $\mathcal{S}(x,t)$ is
a complex-valued phase depending on the position and time, the TDSE for
a particle of mass $m$ in a potential $V$ acquires the form of a
Hamilton-Jacobi equation,
\be
 \frac{\partial \mathcal{S}}{\partial t}
  + \frac{(\nabla \mathcal{S})^2}{2m}
  + V - i\hbar \ \! \frac{\nabla^2 \mathcal{S}}{2m} = 0 .
 \label{e2}
\ee
This is the CQHJ equation, where the last term plays the role of a
nonlocal, complex quantum potential.
Note that, due to the one-to-one correspondence between $\Psi$ and
$\mathcal{S}$ (both functions provide exactly the same information),
Eq.~(\ref{e2}) can also be regarded as the logarithmic form of the
TDSE, since its solution ($\mathcal{S}$) is proportional to $\ln\Psi$.
Equation~(\ref{e2}) can be further generalized by analytic
continuation assuming that both $\mathcal{S}$ and $\Psi$ are
complex-valued functions of a complex (space) variable $z$,
i.e., $\bar{\mathcal{S}} \equiv \mathcal{S}(z,t)$ and
$\bar{\Psi} \equiv \Psi(z,t)$.
Now, analogously to the standard Hamilton-Jacobi formalism, a family
of characteristics or trajectories satisfying the motion law (or
``guidance'' condition) can be defined as
\be
 \bar{v} \equiv \dot{z} = \frac{\nabla \bar{\mathcal{S}}}{m}
  = \frac{\hbar}{im} \ \! \frac{\nabla \bar{\Psi}}{\bar{\Psi}} ,
 \label{e3}
\ee
where $\bar{v}$ is, like $\bar{\mathcal{S}}$ and $\bar{\Psi}$, a
complex-valued, time-dependent field that depends on the (complex)
variable $z$.
Despite this formulation may result inconvenient interpretively, it has
been shown \cite{wyatt11,wyatt12,wyatt13,wyatt21,wyatt22,tannor11,%
tannor12,tannor13,tannor2}, however, that numerical algorithms based
on it are relatively stable and efficient for low-dimensional systems.

Within a real quantum Hamilton-Jacobi (RQHJ) formulation (the standard
Bohmian mechanics) at least two dimensions are necessary in order to
observe quantum vorticality \cite{angel-adsorb1,angel-adsorb2}.
However, as shown below, only one dimension is required to observe the
same phenomenon within the CQHJ framework provided quantum interference
is involved.
The collision of two identical Gaussian wave packets in 1D constitutes
an ideal scenario which illustrates fairly well the appearance of
vorticality in the complex plane.
Before entering into details, first we would like to specify that by
collision of two wave packets (either Gaussian or of any other general
type) here we mean the problem described by a ``one-body'' wave
function which consists of a wave packet superposition.
These wave packets fulfill two conditions initially: (a) they move
towards each other and (b) their respective propagation velocities are
larger than their spreading rates.
With these conditions, after the collision (maximal interference) takes
place, two emerging or outgoing wave packets are clearly defined, just
like in a classical elastic particle-particle scattering problem.
Diffraction-like situations (i.e., those where typical diffraction
patterns can be observed after the collision, instead of two emerging
wave packets) constitute the opposite case.

\begin{figure}
 \includegraphics[width=5cm]{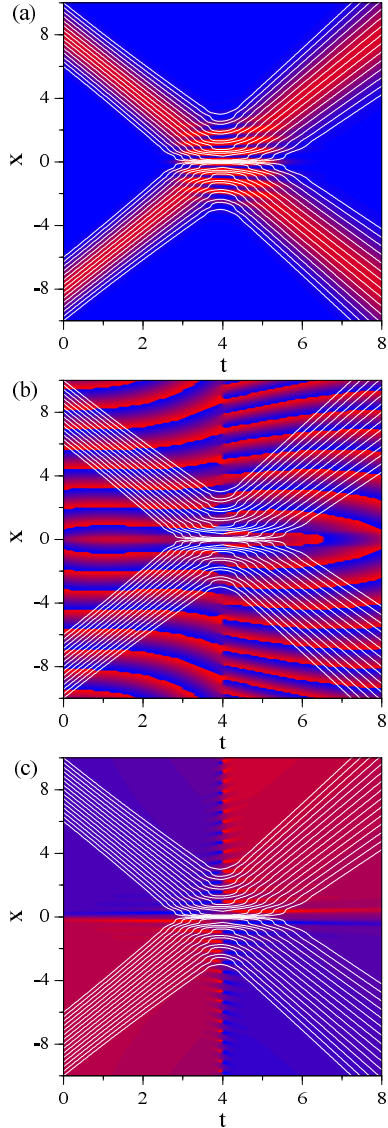}
 \caption{\label{fig1}
  (Color online.)
  Space-time contour plots of the probability density (a), phase (b)
  and velocity (c) fields.
  The associate flux lines or real quantum trajectories (white solid
  lines) have been represented in all plots to make easier their
  understanding.
  The initial values are $\sigma_0 = 1$, $a_j = \mp 8$, and $v_0^{(j)}
  = \pm 2$ (with $j = 1,2$), for a particle with unit mass (arbitrary
  units are used).
  The color scale from red to blue ranges from high values of the
  corresponding field to the lower ones [0 in (a); negative in (b)
  and (c)].}
\end{figure}

The Gaussian wave-packet collision is an analytical problem which,
despite its simplicity, could be considered as representative of other
more complicated, realistic processes characterized by interference
(e.g., scattering problems, diffraction by slits, or quantum control
scenarios).
As indicated above, this process can be described as
\be
 \Psi (x,t) = \mathcal{N} \left[ \psi_1 (x,t) + \psi_2 (x,t) \right] ,
 \label{eint}
\ee
where $\mathcal{N}$ is the normalizing prefactor.
Each wave packet is represented as
\be
 \psi_j (x,t) = \left( \frac{1}{2\pi\tilde{\sigma}_t^2}\right)^{1/4}
  \exp \left[
   - \frac{(x - a_j - v_0^{(j)} t)^2}{4\tilde{\sigma}_t\sigma_0}
   + \frac{i p_0^{(j)} x}{\hbar} - \frac{i E^{(j)} t}{\hbar} \right] ,
 \label{e6}
\ee
where the complex and real time-dependent spreadings are
\begin{subequations}
\be
 \tilde{\sigma}_t = \sigma_0
  \left( 1 + \frac{i\hbar t}{2m\sigma_0^2} \right)
 \label{spread1}
\ee
and
\be
 \sigma_t =\sigma_0
  \sqrt{1 + \left(\frac{\hbar t}{2m\sigma_0^2}\right)^2} ,
 \label{spread2}
\ee
\end{subequations}
respectively, and the initial width (a real-valued quantity)
is given by $\sigma_0$.
Regarding the other parameters, $a_j$ is the initial position of the
center of the wave packet, and $v_0^{(j)} = p_0^{(j)}/m$ and $E^{(j)}$
are the corresponding velocity and energy, respectively .
The space-time contour plots of the probability density, phase and
velocity fields associated with $\Psi$, given by
\begin{subequations}
\begin{eqnarray}
 \rho & = & \Psi^* \Psi ,
 \label{psi1} \\
 S & = & \frac{\hbar}{2i} \ \! \ln \left(\frac{\Psi}{\Psi^*}\right) ,
 \label{psi2} \\
 v & = & \dot{x} = \frac{\nabla S}{m} ,
 \label{psi3}
\end{eqnarray}
\end{subequations}
respectively, are shown in Fig.~\ref{fig1}.
It is worth mentioning that Eqs.~(\ref{psi1}) and (\ref{psi2}) are
transformation relations between the wave field ($\Psi$ and its complex
conjugate) and the flow or hydrodynamic fields ($\rho, S$); the inverse
transformation is just given by the polar form of the wave function,
$\Psi = \rho^{1/2} e^{iS/\hbar}$, and its complex conjugate.
On the other hand, Eq.~(\ref{psi3}) arises after the TDSE is recast in
a RQHJ form,
\be
 \frac{\partial S}{\partial t} + \frac{(\nabla S)^2}{2m} +
  V - \frac{\hbar^2}{2m} \frac{\nabla^2 \rho^{1/2}}{\rho^{1/2}} = 0 ,
 \label{e2bis}
\ee
plus a continuity (or conservation) equation for the probability
density,
\be
 \frac{\partial \rho}{\partial t} +
  \nabla \left( \rho \ \! \frac{\nabla S}{m} \right) = 0 .
 \label{e2tris}
\ee

In order to illustrate the dynamics associated with $\Psi$, several
(real) quantum trajectories $x(t)$ (white solid lines), which are
solutions of Eq.~(\ref{psi3}), have also been added on each plot.
As seen in Fig.~\ref{fig1}(a), the interference of the two wave packets
leads to the appearance of a nodal structure in $\rho$, which makes the
trajectories to avoid certain space regions.
These nodes strongly affect the space-time structures of $S$ and $v$,
as seen in Figs.~\ref{fig1}(b) and \ref{fig1}(c), respectively: sudden
changes from $-\pi$ (blue) to $\pi$ (red) around $t_{\rm max} = 4$
(maximum interference time) in $S$, and a sharp variation from positive
to negative values in $v$.
As clearly noticed from $v$, the configuration (position) space is
divided into two well-defined dynamical regions, where particles will
strictly move with either positive (reddish regions) or negative
(bluish regions) momentum.
Moreover, at $x=0$, there is a sort of ``interface'' acting like a
(fictitious) barrier, which has a determining influence on the topology
of the quantum trajectories: as they start approaching that barrier,
they undergo a strong repulsion and bounce backwards.
Also, around $x=0$, $v$ displays a series of local periodic maxima (in
the regions with $v > 0$) and minima (for $v < 0$), which become more
prominent as time evolves and that, after $t_{\rm max}$, they
interchange their role (i.e., maxima become minima and vice versa).
These structures are connected with the nodes of $\rho$ arising
after the overlapping of both counter-propagating wave packets.
If both wave packets are relatively far apart this effect is so tinny
that it is meaningless dynamically (i.e., it does not affect the
topology of the quantum trajectories), although it does not mean it
does not exist ---it persists because of the initial coherence between
$\psi_1$ and $\psi_2$.

\begin{figure}
 \includegraphics[width=17cm]{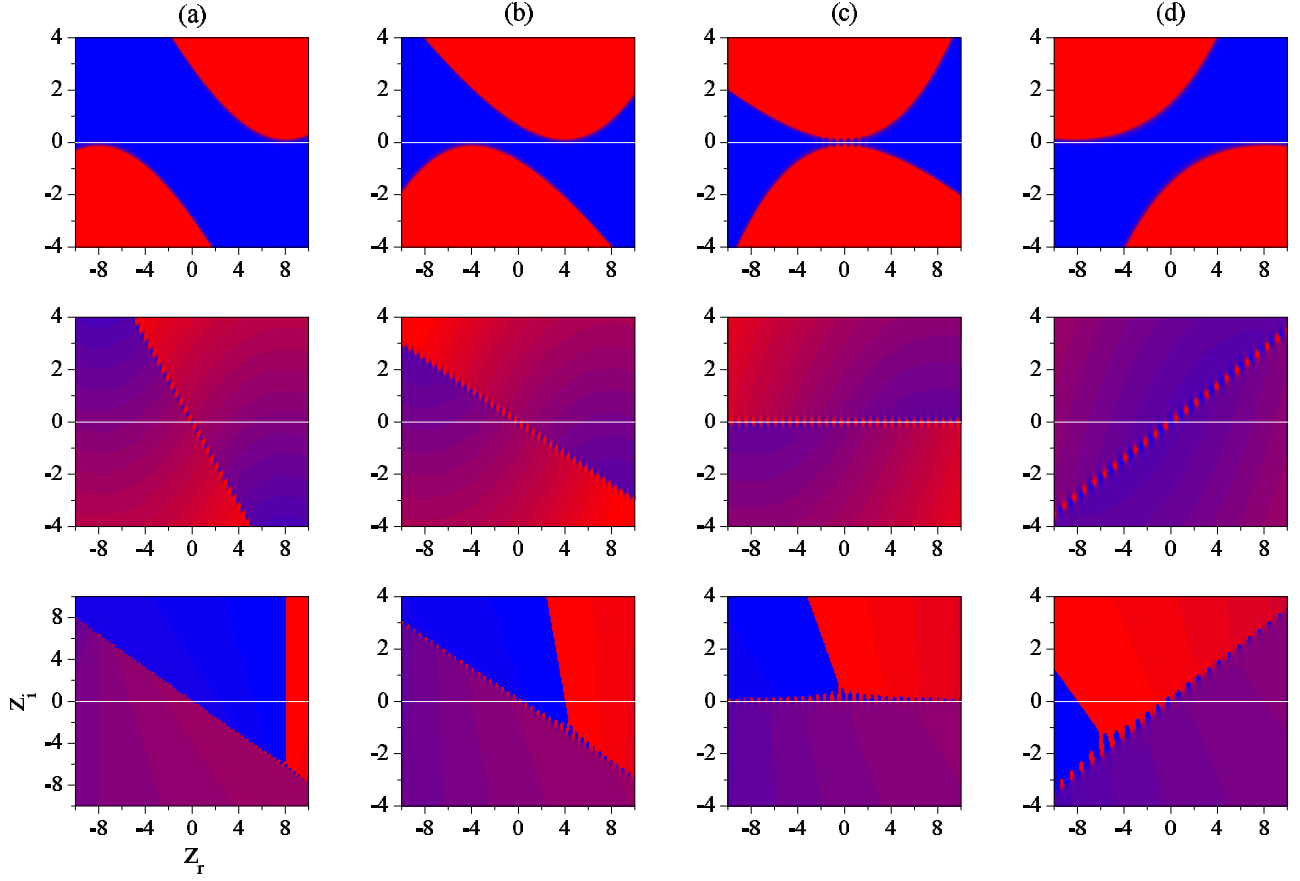}
 \caption{\label{fig2}
  (Color online.)
  The three rows, from top to bottom, correspond to the contour plots
  of $\varrho_{\bar{\Psi}}$, $\varrho_{\bar{v}}$ and $\varphi_{\bar{v}}$,
  respectively, at: (a) $t = 0$, (b) $t = 2$, (c) $t = 4$, and
  (d) $t = 8$ (arbitrary units are used).
  The color scale from red to blue ranges from high values of the
  corresponding field to low ones (0 in the top and middle rows,
  and negative in the bottom one).
  The real axis ($z_i = 0$) is denoted by a white solid line in all
  plots.}
\end{figure}

In the complex version of the wave packet interference process, the
dynamics becomes richer: this 1D problem unfolds into a 2D one on the
complex or {\it Argand} plane, with the dynamics exhibiting more
intricate features.
Here, we are dealing with complex fields (the wave function and the
velocity) which are functions of a complex variable and time.
In order to provide a clear picture of the time-evolution of these
fields, we will decompose both of them in polar form, i.e.,
\be
 \mathcal{F}(z,t) = \varrho_F(z,t) e^{i\varphi_F(z,t)} ,
 \label{edecomp}
\ee
where $\varrho_{\mathcal{F}}(z,t)$ and $\varphi_{\mathcal{F}}(z,t)$
represent, respectively, the modulus and the phase of the complex field
$\mathcal{F}(z,t)$ ---in our case, $\mathcal{F}$ stands for
${\bar {\Psi}}$ and $\bar{v}$.
Thus, in Fig.~\ref{fig2} the contour plots of $\varrho_{\bar{\Psi}}$,
$\varrho_{\bar{v}}$ and $\varphi_{\bar{v}}$ are displayed at four different
times to illustrate the dynamical evolution in the complex plane.
We have not plotted the field $\varphi_{\bar{\Psi}}$ because it is
highly oscillating in the space (i.e., on the Argand plane) and time
ranges considered, and therefore, very difficult to visualize; instead,
we have shown the fields $\varrho_{\bar{v}}$ and $\varphi_{\bar{v}}$, which are
related and provide a more clear information.
Several remarks are worth stressing.
First, as can be inferred from the sequence presented in the upper
row of Fig.~\ref{fig2}, $\Psi(x,t)$ corresponds to the value of
$\bar{\Psi}(z,t)$ along the real axis ($z_r=x$, $z_i=0$) at the
time $t$.
Second, $\bar{\Psi}$ satisfies the normalization condition {\it only}
along the real axis, but not in general on the complex plane.
And, third, following the sequence in Fig.~\ref{fig2} (from left to
right), we observe that the interference process translates into a 2D
anticlockwise rotating dynamics, where at $t_{\rm max}$ the nodal
structure ---a set of aligned nodes--- just lies on the real axis.
At any other time, there is still a nodal alignment, but it is out of
the real axis.
This explains why, in real space, interference is weaker at any other
time than $t_{\rm max}$ (in other words, the larger $|t-t_{\rm max}|$,
the weaker the interference pattern).
Conversely, as $\bar{v}$ shows, the nodal structure remains even for
relatively large times ($t \gg t_{\rm max}$) in the complex space.
Taking into account all these observations, we can say that, within
this (complex) formulation, the evolution of (real) $\Psi$ can be
understood as an ``apparent'' effect of the evolution of $\bar{\Psi}$
in the complex plane.
That is, the value displayed by $\Psi$ at each time can be compared
with the frames of a movie tape (which is the role played by
$\bar{\Psi}$); each frame is watched only when the corresponding
piece of the tape is passing in front of the projector.
The sensation of motion then appears when the tape runs in front
of the projector (i.e., many frames passing consecutively).

\begin{figure}
 \includegraphics[width=12cm]{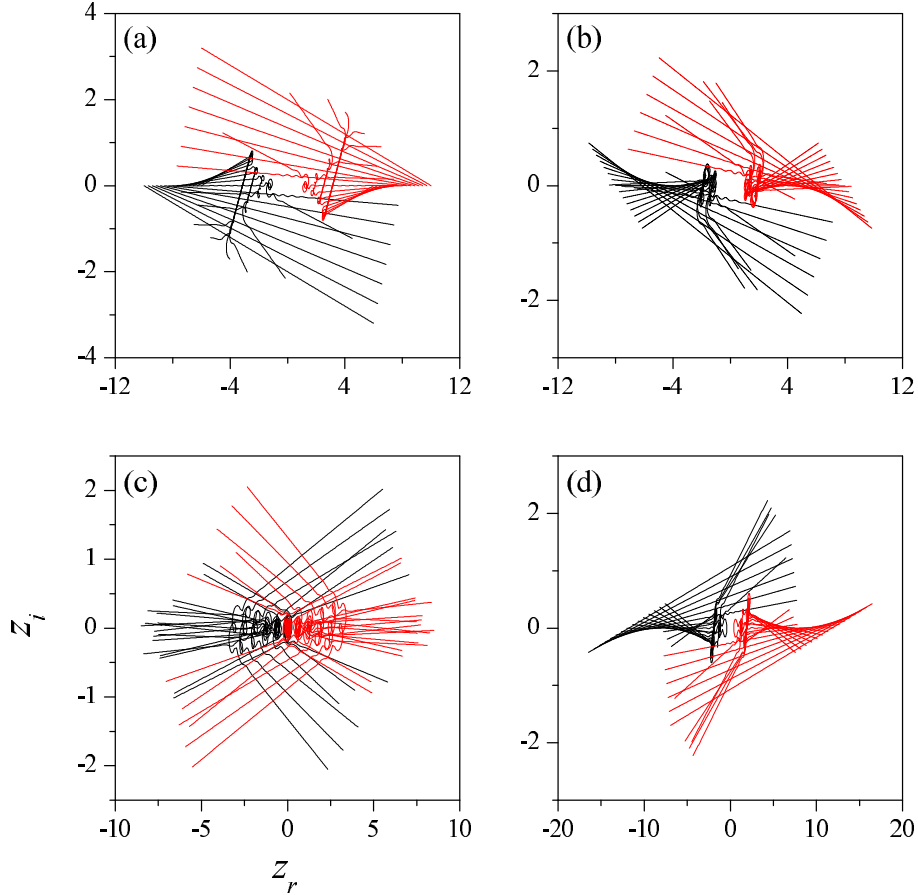}
 \caption{\label{fig3}
  (Color online.)
  Isochrones crossing the real axis [$z_i(t_c) = 0$] at: (a) $t_c=0$,
  (b) $t_c=2$, (c) $t_c=4$, and (d) $t_c=8$, in accordance with the
  snapshots shown in Fig.~\ref{fig2} (arbitrary units are used).
  All the trajectories are propagated from $t=0$ up to $t=8$; the
  crossing points correspond to the same positions reached by the
  real trajectories in Fig.~\ref{fig1} at the corresponding times.
  Black and red trajectories are associated with $\psi_1$ and
  $\psi_2$, respectively.}
\end{figure}

In Fig.~\ref{fig3}, the evolution from $t=0$ to $t=8$ for four
different families of complex trajectories is plotted.
Each family represents a set of {\it isochrones} \cite{wyatt11,%
wyatt12,wyatt13,tannor11,tannor12,tannor13}, i.e., all the trajectories
belonging to the same family cross the real axis (their imaginary part,
$z_i$, vanishes) at a given time, namely the crossing time $t_c$ (in
the cases depicted, at $t_c=0$, 2, 4 and 8, respectively).
Moreover, in our case, the trajectories of each family have been chosen
in such a way that their respective real part coincide with the
positions of the real trajectories in Fig.~\ref{fig1} at the time
they cross the real axis.
Comparing the real and complex trajectory dynamics, it is clear that
there is no a simple one-to-one correspondence between both types of
trajectories, although they are associated with the same physical
problem; real trajectories are not the real part of the complex ones
at any time, as suggested elsewhere \cite{yang1,yang2,yang3}.
To establish a connection, one has to consider the movie-based analogy
between $\Psi$ and $\bar{\Psi}$ pointed out above and the previous
discussion in terms of isochrones.
Accordingly, a single real trajectory is made of the crossings of many
{\it different} complex trajectories with the real axis ---one crossing
for each (real) position at each time.
Note that this allows us to define a real trajectory as a family of
complex trajectories fulfilling the property that their subsequent
crossings (in time) with the real axis generate such a real trajectory.
This is, precisely, the reason why when using computational methods
based on complex trajectories one needs to consider isochrones to
reproduce the corresponding observable \cite{wyatt11,wyatt12,wyatt13,%
tannor11,tannor12,tannor13}.
As seen in Fig.~\ref{fig3}(c), some of these isochrones can display the
effects of a vortical dynamics, unlike the analogous situation in real
configuration space, where vorticality can only be observed in two (or
higher) dimensions \cite{angel-adsorb1,angel-adsorb2}.
Nevertheless, the presence of vortices in complex space can be
explained as in a real dynamical framework.
In the latter, they appear as a consequence of the presence of nodes in
the wave function \cite{angel-adsorb1,angel-adsorb2}.
In such a case, the rotational of the velocity field does not vanish
and trajectories undergo loops around the nodes for some time.
Moreover, the motion along closed loops is shown to be quantized,
i.e., the circle integral of the action evaluated along a trajectory
enclosing a node has a finite, quantized value proportional to the
total number of closed loops around the node.
Going now to the complex framework, we know that except for a constant
in its phase $\varphi_{\bar{\Psi}}$, $\bar {\Psi}$ is uniquely
determined, i.e., it remains invariant under a change of phase provided
$\varphi_{\bar{\Psi}}' = \varphi_{\bar{\Psi}} + 2n\pi \hbar$, with $n$
being an integer number.
Since $\bar{\Psi}$ is a smooth function, discontinuities in its phase
($n \neq 0$) can only occur in nodal regions, where the wave function
vanishes and the phase displays ``jumps'' due to its multivaluedness.
These discontinuities give rise to a vortical dynamics, as infers from
the circulation of the phase $\varphi_{\bar{\Psi}}$ along a closed path:
when $n \neq 0$, $\bar{v}$ is rotational, this leading to the formation
of quantum vortices around the nodal regions.
The appearance of this dynamics breaks off the causticity regime
associated with free wave packet propagation, where (complex) quantum
trajectories give rise to the appearance of {\it caustics}, i.e.,
curves arising as the envelope of a set of trajectories (all of them
tangent to such a curve at different, consecutive times).
This can be seen in panels (a), (b) and (d) of Fig.~\ref{fig3}.
Before and after $t_{\rm max}$, the nearly free propagation of $\psi_1$
and $\psi_2$ manifests as a sort of causticity regime, which can not be
appreciated at all under a strong vortical dynamics, as seen in
Fig.~\ref{fig3}(c), where this dynamics prevents the isochrones
to display the corresponding caustics.

To conclude, from a theoretical and interpretative viewpoint, we
have shown that very intricate and rich, complex dynamics can appear
provided interference is present even in the case of very simple
processes (with simple dynamics in real configuration space).
These dynamical behaviors deserve much attention in the design and
improvement of numerical techniques based on the CQHJ formalism, since
the interplay of vorticality and causticity might become relevant
sources of inefficiency when dealing with realistic problems.
The knowledge of the complexity involved by the vortical dynamics
should be therefore taken into account in the construction and
implementation of numerical methods aimed to describe more realistic
situations.
We would like to note that, rather than being independent, the two
quantum singularities (vortices and caustics) are related through the
quantum nonlocality \cite{sanz-cpl}.
In this sense, an immediate, natural extension of the study presented
here would be multi-interference phenomena, such as the {\it Talbot
effect} \cite{talbot}, which are currently being developed and that
constitute an intermediate step before going to more complicated
situations, such as surface scattering.
It is expected that the information obtained from this type of studies
will shed some light on whether working within the CQHJ framework can
be further developed and applied to higher-dimensional systems.
Up to date (real) trajectory-based methods have been proven to be
valuable numerical and interpretative tools to explore this kind of
quantum problems; CQHJ methods would further benefit from the
advantages of working in the complex space, which have led to
well-known numerical simplifications in problems of interest in many
other fields of Physics.
In particular, the problem we have tackled is a special case due
to the dynamical richness mentioned above.
However, when dealing with problems with two o higher dimensions, the
computational effort of complex trajectories is significantly
increased since the dimensionality of the working space is double.

This work has been supported by DGCYT (Spain) under project with
reference FIS2007-62006.
A.S.\ Sanz would also like to acknowledge the Spanish Ministry of
Education and Science for a ``Juan de la Cierva'' Contract.


\end{document}